# Giant magnetocaloric effect in $Gd_2NiMnO_6$ and $Gd_2CoMnO_6$ ferromagnetic insulators

J. Krishna Murthy[1], K. Devi Chandrasekhar[2], Sudipta Mahana[3], D. Topwal[3], and A. Venimadhav[1]

[1]*Cryogenic Engineering Centre, Indian Institute of Technology, Kharagpur-721302, India*
[2]*Department of Physics and Center for Nanoscience and Nanotechnology, National Sun-Yat Sen University, Kaohsiung 804, Taiwan*
[3]*Institute of Physics, Bhubaneswar-751005, India.*

## Abstract

We have investigated magnetocaloric effect in double perovskite $Gd_2NiMnO_6$ (GNMO) and $Gd_2CoMnO_6$ (GCMO) samples by magnetic and heat capacity measurements. Ferromagnetic ordering is observed at ~130 K (~112 K) in GNMO (GCMO), while the Gd exchange interactions seem to dominate for T < 20 K. In GCMO, below 50 K, an antiferromagnetic behaviour due to 3d-4f exchnage interaction is observed. A maximum entropy ($-\Delta S_M$) and adiabatic temperature change of ~35.5 J Kg$^{-1}$ K$^{-1}$ (~24 J Kg$^{-1}$ K$^{-1}$) and 10.5 K (6.5 K) is observed in GNMO (GCMO) for a magnetic field change of 7 T at low temperatures. Absence of magnetic and thermal hysteresis and their insulating nature make them promising for low temperature magnetic refrigeration.

--------------------------------------------------------

Email of corrosponding author: venimadhav@hijli.iitkgp.ernet.in

## 1. Introduction

Magnetic refrigerant materials exhibit a temperature change when exposed to adiabatic demagnetization. The phenomena is known as solid-state magnetic refrigeration [1, 2] and strong coupling of the magnetic moment with lattice thermal energy is desirable in such materials. Magnetic refrigeration is an emerging technology alternative to the conventional gas-compression refrigeration in food preservation and air conditioning applications. Importantly, solid state cooling offers noise-free and energy efficient refrigeration suitable for room temperature cooling and cooling of microelectronic components [2-4]. Solid state cooling is essential for the growing needs of low temperature applications in space, particle detectors, and medical applications [2, 3, 5-7]. Liquefaction of $H_2$ (20 K) by magnetocaloric method has been reported to be cost effective and could render hydrogen to be a competitive alternative fuel [8]. To achieve cooling below 1 K, the adiabatic demagnetization refrigeration is an attractive process compared to $^3$He/$^4$He dilution refrigeration because of the growing cost of helium and scarcely available $^3$He isotope.

For low temperature magnetic refrigerant materials, it is important to have: (i) large effective spin quantum number, (ii) low magnetic anisotropy and low magnetic ordering temperature, (iii) small specific heat (iv) large magnetization under magnetic field, and (v) weak magnetic exchange interactions [9]. In magnetic refrigeration technology, Gd and Gd based alloys [10, 11], large molecular materials [12] have continued to receive large attention because large magnetic moment of Gd. Recently, $R_2BMnO_6$ (R= rare earth, B= Ni, Co) ceramics have attracted extensive research attention due to their ferromagnetic insulating behaviour with large magneto-dielectric effect and ferroelectric characteristics [13-15]. Coexistence of magnetocaloric and electrocaloric effects with multicaloric coupling in $Y_2CoMnO_6$ has excited the role of double perovskites in solid state refrigeration applications [16]. In this study we present the occurrence of large magnetic entropy change ~ 35.5 J Kg$^{-1}$ K$^{-1}$ and ~ 24 J Kg$^{-1}$ K$^{-1}$ in double perovskite GNMO and GCMO samples respectively; from heat capacity, a large adiabatic temperature change of ~ 10.5 K at low temperature has been noticed in GNMO. These values are much larger than the parent $La_2NiMnO_6$ and $La_2CoMnO_6$ compounds [17, 18].

## 2. Experimental details:

Polycrystalline $Gd_2BMnO_6$ (B= Ni and Co) samples were prepared by solid-state reaction method using high purity precursor materials of $Gd_2O_3$, $NiO$, $Co_3O_4$ and $MnO_2$. Stoichiometric proportion of the powders were thoroughly mixed and heated at temperatures from 1000°C to 1350°C with an intermediate grindings until the desired phase was obtained. The structural analysis was done using a Phillips powder x-ray diffractometer with Cu-K$_\alpha$ radiation. Rietveld refinement has revealed monoclinic crystal structure of P2$_1$/n space group for both the samples similar to other double perovskite compounds [13-15, 19]. Magnetic measurements were done with an EverCool Quantum Design SQUID-VSM magnetometer. The heat capacity at constant magnetic fields was measured using relaxation method in a Quantum design-physical property measurement system (PPMS).

## 3. Results and discussion

**Fig. 1(a) & (b)** shows the temperature dependent magnetization, M (T) in zero-field cooled (ZFC) and field-cooled (FC) modes with 0.01 T DC field for both the samples. In GNMO, a paramagnetic (PM) to ferromagnetic (FM) ordering is observed at $T_C \sim 132$ K and a rise in $M_{FC}$ at low temperatures (T < 20 K) can be attributed to the polarization of Gd spins with $Gd^{3+}$-$O^{2-}$-$Gd^{3+}$ exchange interactions. In case of GCMO, PM to FM phase transition is noticed at $T_C \sim 112$ K and it is followed by a broad peak around ~ 47 K suggesting the polarization of magnetic moments with an antiferromagnetic (AFM) like behaviour. At low temperatures, an increasing magnetization due to Gd can be found similar to GNMO. The FM ordering can be attributed to the super exchange interaction of $Ni^{2+}$-O-$Mn^{4+}$ ($Co^{2+}$-O-$Mn^{4+}$) magnetic species in GNMO (GCMO). The AFM arrangement of magnetic moments at ~ 47 K in GCMO can be ascribed to the negative 3d-4f exchange interactions in between FM network of Co/Mn sublattice and Gd spins, like in $RMnO_3$ and $RCrO_3$ (R= Gd, Ho, and Dy) systems [20-23]. We have observed no bifurcation in between field-cooled cooling and field-cooled warming modes of magnetization (not shown here) that suggests the absence of thermal hysteresis in both the samples.

Temperature dependent magnetic susceptibility in PM state is fitted to Curie-Weiss (CW) law with an additional susceptibility contribution of Gd magnetic moments and is vgiven as,

$$\chi = \frac{C_{(Ni/Co)-Mn}}{T-\theta} + \frac{C_{Gd}}{T} \qquad\qquad \text{---------} \quad (1)$$

Here, $C_{(Ni/Co)-Mn}$ and $C_{Gd}$ are the Curie constants of (Ni/Co)-O-Mn network and Gd sublattices respectively. Inset to **Fig. 1**(a) & (b) shows the $\chi^{-1}$ vs. T data of GNMO and GCMO samples and the fit to the Eqn. (1) is shown by the solid line. The Curie constant is estimated using the formula,

$$\mu_{\text{eff}} = \frac{1}{\mu_B}\left(\frac{3K_B\, C_M}{N_A}\right)^{\frac{1}{2}}$$

Where, $K_B$ is the Boltzmann constant and $\mu_B$ is Bohr magneton. From the fitting, in GNMO sample, the obtained CW temperature ($\theta$) = 145 K and the effective PM moments $(\mu_{\text{eff}})_{\text{Ni-Mn}}$ and $(\mu_{\text{eff}})_{\text{Gd}}$ are ~ 6.37 $\mu_B$ and ~ 13.2 $\mu_B$ respectively. In GCMO, $\theta$ ~ 111.2 K and $(\mu_{\text{eff}})_{\text{Co-Mn}}$ ~ 5.81 $\mu_B$ and $(\mu_{\text{eff}})_{\text{Gd}}$ ~ 14.49 $\mu_B$. In both the samples, the effective PM moment values are close to the theoretically calculated spin only contribution of (Ni/Co)-O-Mn network and rare-earth Gd spins.

Thermal evolution of heat capacity ($C_P$) of GNMO and GCMO samples under different magnetic fields (i.e., 0, 1, 3, 5 and 7 T) are depicted in **Fig.1**(c) & (d) respectively. The $C_P$ vs. T displays distinct transitions in both the samples. A λ-type anomaly under zero field at ~130 K in GNMO and ~112 K in GCMO samples correspond to the second-order magnetic phase transition. In both the samples, with the increase of magnetic field, magnitude of the peak decreases and shifts to high-temperature side which is the characteristic feature of FM ordering. The broadening of $C_P$ (T) peak under external field is due to the randomization of magnetic moments in wide temperature region. Above 20 K, $C_P$(T) increases with increasing temperature and follows the $T^3$ dependence due to the lattice contribution [24], and all $C_P$ vs. T curves gets merged for different magnetic fields as shown in the inset of **Fig. 1** (c) & (d). The increasing trend in heat capacity below ~ 20 K can be noticed in both GNMO and GCMO samples and can be attributed to the Gd magnetic contribution. With the application of magnetic field, $C_P$(T) value increases and broadens the dip at ~20 K and shifts to higher temperature; this anomalous behaviour can be attributed to the schottkey contribution that arises from the splitting of degenerate ground state energy levels at the $Gd^{3+}$ state in crystal fields [25, 26]. In contrast to GNMO sample, a peak in $C_P$ (T) can be noticed at ~5 K in GCMO corresponding to the onset of AFM ordering of Gd magnetic moments. In GNMO, this peak can be noticed only with the magnetic fied and this suggests the Gd ordering occurs at temperatures lower than 2 K.

Isothermal field-dependent magnetization i.e., M (H) of GNMO and GCMO samples at 2 K is shown in the **Fig. 2(a)**. Both the samples show no hysteresis and such kind of magnetic reversibility in M (H) is beneficial for the solid-state magnetic refrigeration. GNMO shows

S-shaped M(H) behaviour with significant changes in magnetization at low fields and saturation like behaviour for H $\geq$ 6 T. The observed saturation magnetization ($M_S$) value of ~18.9 $\mu_B$/f.u. matches well with the theoretically estimated value of 19 $\mu_B$/f.u (5 $\mu_B$/f.u for $Ni^{2+}$-O-$Mn^{4+}$ pair and 14 $\mu_B$/f.u for $Gd_2$). While GCMO exhibits a linear variation of magnetization from low fields to 4 T (see inset of Fig 2(a)) and shows no saturation even for H $\geq$ 6 T. Further, in GCMO, the magnetization value at high field (7 T in our set up) is ~16.5 $\mu_B$/f.u., which is smaller than theoretically estimated sum of the fully polarized $Co^{2+}$-O-$Mn^{4+}$ interaction (6 $\mu_B$/f.u.) and magnetic moment of Gd spins (14 $\mu_B$/$Gd_2$). Incomplete saturation of magnetization in GCMO can be related to the presence of significant 3d-4f negative AFM exchange correlations. **Fig**. **2**(b) & (c) shows the representative isothermal M (H) plots of GNMO and GCMO samples taken in the temperature range of 2-40 K. With this data, we have calculated MCE using Maxwell's relation [27],

$$\Delta S_M (T, H) = \int_0^H \left(\frac{\partial M}{\partial T}\right)_H dH$$

Since, isothermal M (H) curves are measured by discrete field changes, the following expression is used,

$$-\Delta S_M = \sum_i \frac{M_i - M_{i+1}}{T_{i+1} - T_i} \Delta H_i, \qquad \text{---------- (2)}$$

Here, $M_i$ and $M_{i+1}$ are initial magnetization values at $T_i$ and $T_{i+1}$ respectively for a field change of $\Delta H_i$. In this method, the magnetic entropy change corresponding to the average temperature T (= ($T_1$+$T_2$)/2) is given by the area enclosed by two consecutive isothermal M (H) curves at $T_1$ and $T_2$ divided by $\Delta T = T_2-T_1$ ($T_2 > T_1$). We have calculated magnetic entropy change (-$\Delta S_M$) using Eqn. (2) and is plotted with temperature for different magnetic fields as shown in the **Fig**. **3**(a) and (b) for GNMO and GCMO samples respectively. The value of -$\Delta S_M$ is positive in the entire temperature region and increases with the magnetic field; this indicates that the magnetic field favors FM ordering. In GNMO, -$\Delta S_M$ increases with the decrease of temperature and a maximum change of entropy of ~ 35.5 J $Kg^{-1}$ $K^{-1}$ is observed at 2 K. In GCMO, a peak in -$\Delta S_M$ can be noticed at ~ 5 K that corresponds to the AFM ordering of Gd and it is consistent with the heat capacity data in **Fig. 1**(d). The maximum change of entropy -$\Delta S^{max}_M$ (peak maximum at 4 K) is ~24 J $Kg^{-1}$ $K^{-1}$ for field change ($\Delta H$) of 7 T. Theoretically, entropy is associated with the magnetic degrees of freedom and can be calculated using $\Delta S_M$ = Rln (2S+1), where R is the universal gas constant and S is the total spin quantum number and they are 40.68 J $Kg^{-1}K^{-1}$ and 41.84 J $Kg^{-1}K^{-1}$ for GNMO and GCMO respectively. Since the lattice contribution to the total entropy is negligible at low temperature, the maximum value of MCE calculated at 2 K is 82 % in GNMO and 56 % in GCMO with respect to the theoretically estimated magnetic entropy. Low value of $-\Delta S^{max}_{M}$ in GCMO is attributed to the incomplete saturation and low value of magnetization as noticed from M (H) (**Fig**. **2**(a)). Moreover, the negative exchange coupling between (Co-O-Mn) and Gd sublattices can also be an effective cause for the low value of MCE. The MCE values of GNMO are larger than the paramagnetic salts [28, 29] and $HoMn_2O_5$ single crystal [30], while it is comparable to the Gadolinium Gallium Garnets $Gd_3(Ga_{1-x}Fe_x)_5O_{12}$ [24], Gd and 3d-transition metal based small molecular magnetic systems [31], rare-earth manganites ($RMnO_3$, R= Ho, Tb, Gd, Dy and Yb) [32, 33], and magnetically frustrated $EuHo_2O_4$, $EuDy_2O_4$ compounds [34]. Interestingly all these systems have similar magnetic ordering below the liquid hydrogen temperatures (~ 20 K). MCE has also been calculated from the heat capacity measurements using the following thermodynamic relation,

$$\Delta S_M (T, H) = \int_0^T \frac{(C_P(H) - C_P(0))}{T} dT \qquad (or)$$

$$\Delta S_M (T, H) = \int_0^T \frac{C_P(0)}{T} dT, \qquad \text{---------- (3)}$$

Here, $C_P$ (H) and $C_P$ (0) are the heat capacity values measured with field and without field respectively. The entropy change from heat capacity data is shown in the inset of **Fig**. **3**(a) & (b). Here, maximum values of $-\Delta S_M$ and its behaviour calculated from complementary experimental tools have shown a close resemblance. From the heat capacity measurement, a small change in entropy has also been found near to FM-PM ($T_C$) transition in both the samples; $-\Delta S_M$ ~ 3 J Kg$^{-1}$K$^{-1}$ and ~ 2 J Kg$^{-1}$K$^{-1}$ in GNMO and GCMO respectively**.**

Apart from change in entropy, the total adiabatic temperature change, $\Delta T_{ad}$ associated with external magnetic field is an another important parameter for evaluating MCE materials and it is calculated from the temperature dependence of the total entropy[11]. The zero magnetic field corresponding to total entropy *S* (0, T) can be calculated by using heat capacity data as,

$$dS = (C_p dT)/T \quad or$$

$$S(0,T) = \int_0^T \frac{C_p(0,T)}{T} dT \qquad \text{------------- (4)}$$

and magnetic field induced total entropy, *S* (H, T) is calculated by subtracting the corresponding $-\Delta S_M$ from *S* (0, T). Then, $\Delta T_{ad}$ value is estimated from the isoentropic difference in between the entropy curves *S* (0, T) and *S* (H, T). **Fig. 4** shows the temperature dependent $\Delta T_{ad}$ under different magnetic fields for GNMO and GCMO (inset) samples. $\Delta T_{ad}$ shows a peak near FM transition with a magnitude of ~ 1.8 K in GNMO and ~ 0.8 K in GCMO samples for a field change of 7 T. At low temperatures < 20 K with the onset of Gd ordering $\Delta T_{ad}$ shows a peak value of ~ 10.5 K in GNMO and ~ 6.5 K for GCMO samples. We have measured the electrical resistivity in these samples, and it is of the order of $10^6$ $\Omega$m at 150 K and it increases further with the decrease of temperature. Particularly, for low temperature refrigeration, high electric resistivity is desirable as the low resistivity of the materials can induce significant eddy current loss that limits the cooling efficiency of magnetic refrigeration process [9, 35].

## 4. Conclusions

In summary, we have prepared GNMO and GCMO double perovskites by simple solid-state reaction method and studied their magnetocaloric properties. Magnetic and heat capacity measurements on GNMO sample has revealed a superior magnetocaloric performance of -$\Delta S_M$ ~ 35.5 J g$^{-1}$K$^{-1}$ and $\Delta T_{ad}$ ~ 10.5 K at low temperatures compared to GCMO where -$\Delta S_M$ ~ 24 JKg$^{-1}$ K$^{-1}$ and $\Delta T_{ad}$ ~ 6.5 K. Presence of 3d-4f interactions reduce the resultant magnetocaloric effect in GCMO. Further, simple synthesis, high chemical stability, absence of magnetic and thermal hysteresis and insulating nature suggest them as potential magnetic refrigerants below the liquid hydrogen temperatures.

**Acknowledgments:**
The authors acknowledge DST and IIT Kharagpur for funding VSM-SQUID magnetometer. Krishna thanks to CSIR-UGC, Delhi for SRF.

REFERENCES

[1] Tishin A M and Spichkin Y I, *The Magnetocaloric Effect and its Applications* (Taylor & Francis) (2003).

[2] GschneidnerJr K A, Pecharsky V K and Tsokol A O 2005 *Rep. Prog. Phys*. **68**, 1479

[3] Yu B F, Gao Q, Zhang B, Meng X Z and Chen Z 2003 *Int. J. Refrigeration* **26**, 622

[4] Franco V, Blázquez J S, Ingale B and Conde A 2012 *Ann. Rev. Mater. Res.* **42**, 305

[5] Hagmann C and Richards P L 1995 *Cryogenics* **35**, 303

[6] Kushino A, Aoki Y, Yamasaki N Y, Namiki T, Ishisaki Y, Matsuda T D, Ohashi T, Mitsuda K and Yazawa T 2001 *J. Appl. Phys.* **90**, 5812

[7] Tishin A M and Spichkin Y I 2014 *Int. J. Refrigeration* **37** 223.

[8] Zhang L, Sherif S A, DeGregoria A J, Zimm C B and Veziroglu T N 2000 *Cryogenics* **40**, 269

[9] Phan M H and Yu S -C 2007 *J. Magn. Man. Mater*. **308**, 325

[10] Dankov S Y, Tishin A M, Pecharsky V K and Gschneidner K A 1998 *Phys. Rev. B* **57**, 3478

[11] Pecharsky V K and Gschneidner J K A 1997 *Phys. Rev. Lett.* **78**, 4494

[12] Sessoli R 2012 *Angew. Chem. Int.l Ed*. **51**, 43.

[13] Krishna Murthy J, Chandrasekhar K D, Murugavel S and Venimadhav A 2015 J. *Mater. Chem. C* **3**, 836

[14] Sharma G, Saha J, Kaushik S D, Siruguri V and Patnaik S 2013 *Appl. Phys. Lett*. **103**, 012903

[15] Yáñez-Vilar S, Mun E D, Zapf V S, Ueland B G, Gardner J S, Thompson J D, Singleton J, Sánchez-Andújar M, Mira J, Biskup N, Señarís-Rodríguez M A and Batista C D 2011 *Phys. Rev. B* **84**, 134427

[16] Krishna Murthy J and Venimadhav A 2014 *J. Phys. D: Appl. Phys.* **47**, 445002

[17] Balli M, Fournier P, Jandl S and Gospodinov M M 2014 *J. Appl. Phys.* **115**, 173904

[18] Balli M, Fournier P, Jandl S, Truong K D and Gospodinov M M 2014 *J. Appl. Phys*. **116**, 073907

[19] Kakarla D C, Jyothinagaram K M, Das A K and Adyam V 2014 *J. Am. Ceram. Soc.,* **97**, 2858

[20] Lin L, Li L, Yan Z B, Tao Y M, Dong S and Liu J M 2013 *Appl. Phys. A* **112**, 947

[21] Lee N, Choi Y J, Ramazanoglu M, Ratcliff W, Kiryukhin V and Cheong S W 2011 *Phys. Rev. B* **84**, 020101

[22]     Zhang N, Guo Y Y, Lin L, Dong S, Yan Z B, Li X G and Liu J-M 2011 *Appl. Phys. Lett* **99**, 102509

[23]     McDannald A, Kuna L and Jain M 2013 *J. Appl. Phys.* **114**, 113904

[24]     Koichi M, Ayumi M, Koji K and Takenori N 2009 *J. Jpn. Appl. Phys.* **48**, 113002

[25]     Köhler U, Demchyna R, Paschen S, Schwarz U and Steglich F 2006 *Physica B: Condens. Matt.* **378**, 263

[26]     Pedro S S, Tedesco J C G, Yokaichiya F, Brandão P, Gomes A M, Landsgesell S, Pires M J M, Sosman L P, Mansanares A M, Reis M S and Bordallo H N 2014 *Phys. Rev. B* **90**, 064407

[27]     Pecharsky V K and Gschneidner K A 1999 *J. Appl. Phys.* **86**, 565

[28]     Fernández A, Bohigas X, Tejada J, Sulyanova E A, Buchinskaya I I and Sobolev B P 2007 *Mater. Chem. Phys.* **105**, 62

[29]     Warburg E 1881 *Annalen der Physik* **249**, 141

[30]     Balli M, Jandl S, Fournier P and Gospodinov M M 2014 *Appl. Phys. Lett.* **104**, 232402

[31]     Zheng Y -Z, Zhou G -J, Zheng Z and Winpenny R E P 2014 *Chem. Soc. Rev.* **43**, 1462

[32]     Midya A, Das S N, Mandal P, Pandya S and Ganesan V 2011 *Phys. Rev. B* **84**, 235127

[33]     Aditya A Wagh, Suresh K G, Anil Kumar P S and Suja Elizabeth 2015 *J. Phys. D: Appl. Phys.* **48**, 135001

[34]     Midya A, Khan N, Bhoi D and Mandal P 2012 *Appl. Phys. Lett.* **101**, 132415

[35]     Shen B G, Sun J R, Hu F X, Zhang H W and Cheng Z H 2009 *Adv. Mater.* **21**, 4545

**Figure Captions:**

**Fig. 1**: M vs. T (K) data in ZFC (solid symbol), FC (open symbol) modes for 0.01 T field for (a) GNMO and (b) GCMO; and their inset shows the CW law fit to the experimental data of $\chi^{-1}$ vs. T (K), (c) & (d) shows the $C_p$ vs. T (K) data for different magnetic fields in GCMO and GNMO samples respectively, and their inset displays the magnified view of $C_P$ vs. T (K) data at T < 40 K.

**Fig. 2**: (a) Isothermal M (H) loops at 2 K for GCMO and GNMO samples, inset shows the blown up portion of the hysteresis loop at low field region, (b) and (c) are representative M (H) isotherms at selective temperatures for GNMO and GCMO samples respectively.

**Fig. 3**: -$\Delta S_M$ vs. T (K) curves measured from the isothermal magnetization curves in (a) GNMO, (b) GCMO samples and their inset shows -$\Delta S_M$ vs. T (K) data measured from the heat capacity.

**Fig. 4**: Temperature dependent $\Delta T_{ad}$ vs. T (K) data in GCMO sample (inset for GNMO) for different magnetic fields.

**Fig. 1**

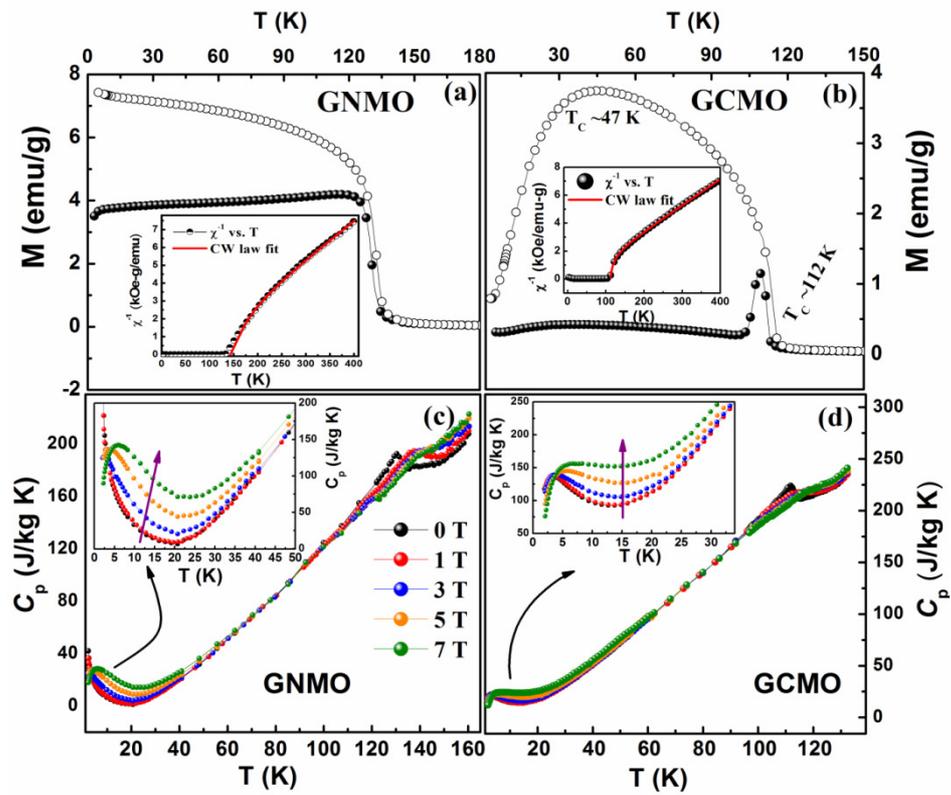

**Fig. 2**

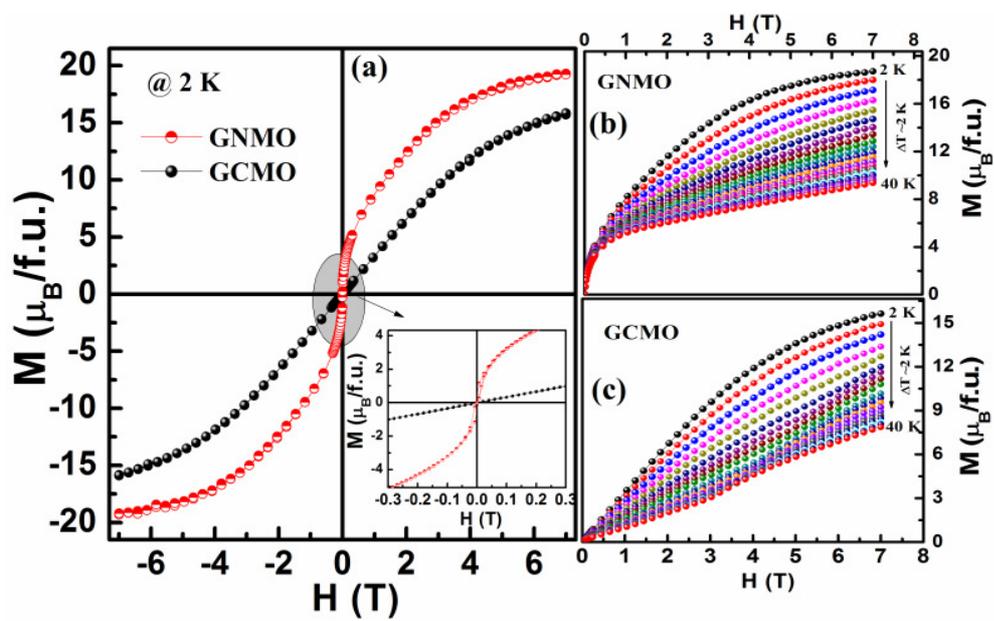

**Fig. 3**

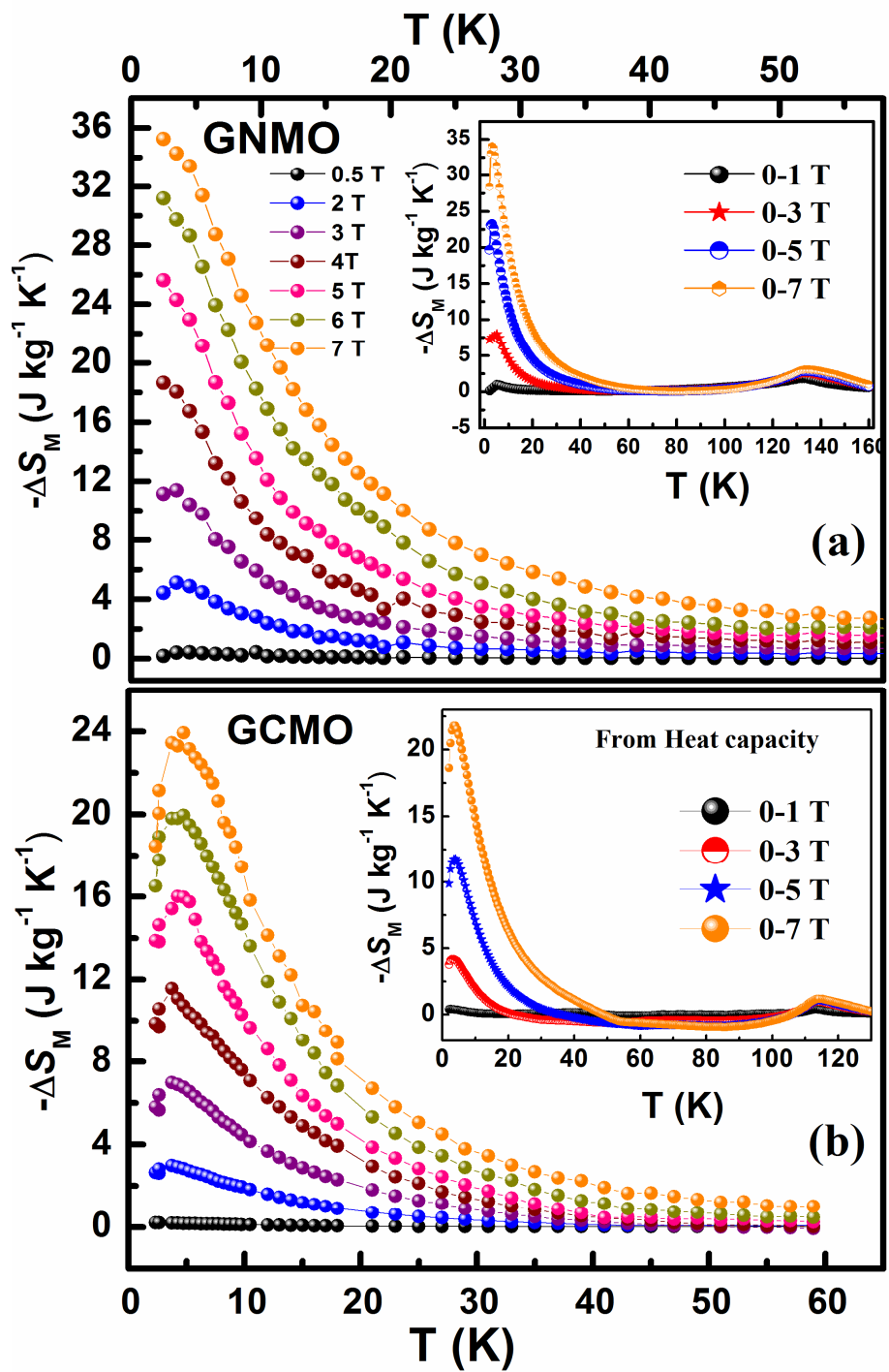

**Fig. 4**

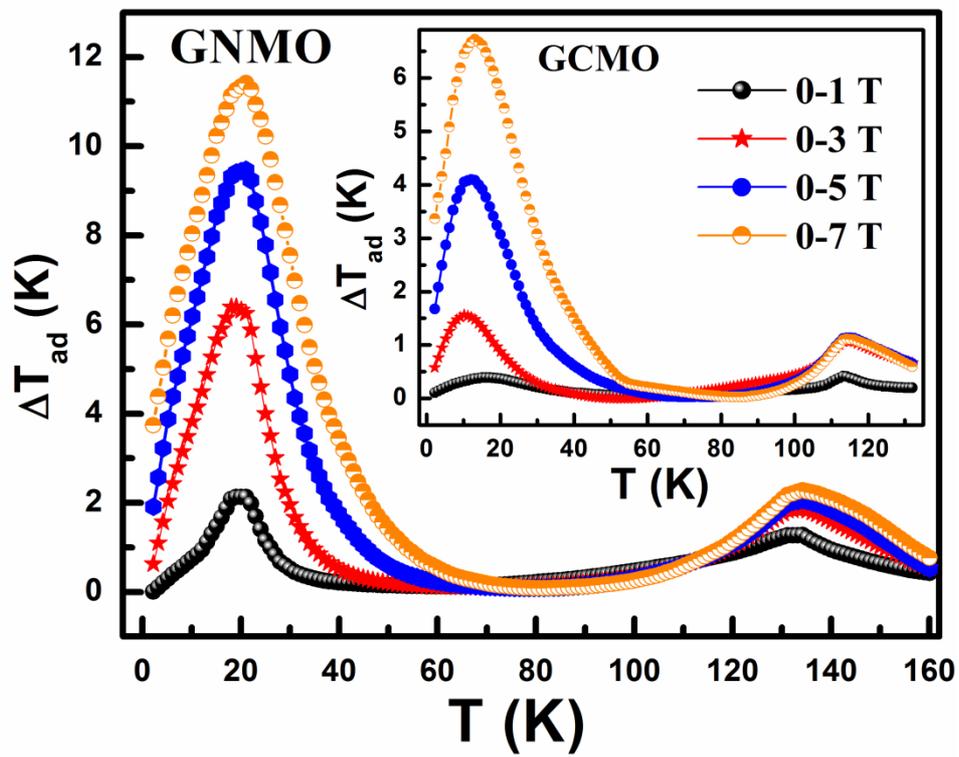